\documentclass[journal]{vgtc}                     

\onlineid{1681}

\vgtccategory{Research}

\vgtcpapertype{application/design study}

\newcommand{\name}{\textit{CSLens}}
\newcommand{\zyt}{\textcolor{black}}
\newcommand{\xlw}{\textcolor{black}}
\newcommand{\revised}{\textcolor{black}}

\newcommand{\controlpanel}{{\textcolor{black}{Control Panel}}}
\newcommand{\temporaloverview}{{\textcolor{black}{Temporal Overview}}}
\newcommand{\chargingstationinfo}{{\textcolor{black}{Charging Station Info}}}
\newcommand{\mapview}{{\textcolor{black}{Map View}}}
\newcommand{\resultview}{{\textcolor{black}{Result View}}}
\newcommand{\detailview}{{\textcolor{black}{Impact View}}}

\title{{\name}: Towards Better Deploying Charging Stations via \\ Visual Analytics —— A Coupled Networks Perspective}

\author{%
  \authororcid{Yutian Zhang}{0009-0002-6711-4478},
  \authororcid{Liwen Xu}{0009-0009-2124-8172},
  \authororcid{Shaocong Tao}{0009-0003-1065-6943},
  \authororcid{Quanxue Guan}{0000-0002-1379-620X},
  \authororcid{Quan Li}{0000-0003-2249-0728},
  and
  \authororcid{Haipeng Zeng}{0000-0002-0339-0361}
}

\authorfooter{
  \item
  Y. Zhang, L. Xu, S. Tao and Q. Guan are with Sun Yat-sen University. E-mail: \{zhangyt85, xulw8, taoshc\}@mail2.sysu.edu.cn, and guanqx3@mail.sysu.edu.cn.
  \item
  Q. Li is with the School of Information Science and Technology, ShanghaiTech University. E-mail: liquan@shanghaitech.edu.cn.

  \item H. Zeng (corresponding author) is with Sun Yat-sen University. E-mail: zenghp5@mail.sysu.edu.cn.
}

\abstract{In recent years, the global adoption of electric vehicles (EVs) has surged, prompting a corresponding rise in the installation of charging stations. This proliferation has underscored the importance of expediting the deployment of charging infrastructure. Both academia and industry have thus devoted to addressing the charging station location problem (CSLP) to streamline this process. However, prevailing algorithms addressing CSLP are hampered by restrictive assumptions and computational overhead, leading to a dearth of comprehensive evaluations in the spatiotemporal dimensions. Consequently, their practical viability is restricted. Moreover, the placement of charging stations exerts a significant impact on both the road network and the power grid, which necessitates the evaluation of the potential post-deployment impacts on these interconnected networks holistically. In this study, we propose {\name}, a visual analytics system designed to inform charging station deployment decisions through the lens of coupled transportation and power networks. {\name} offers multiple visualizations and interactive features, empowering users to delve into the existing charging station layout, explore alternative deployment solutions, and assess the ensuring impact. To validate the efficacy of {\name}, we conducted two case studies and engaged in interviews with domain experts. Through these efforts, we substantiated the usability and practical utility of {\name} in enhancing the decision-making process surrounding charging station deployment. Our findings underscore {\name}'s potential to serve as a valuable asset in navigating the complexities of charging infrastructure planning.
}


\teaser{
  \centering
  \includegraphics[width=0.96\linewidth]{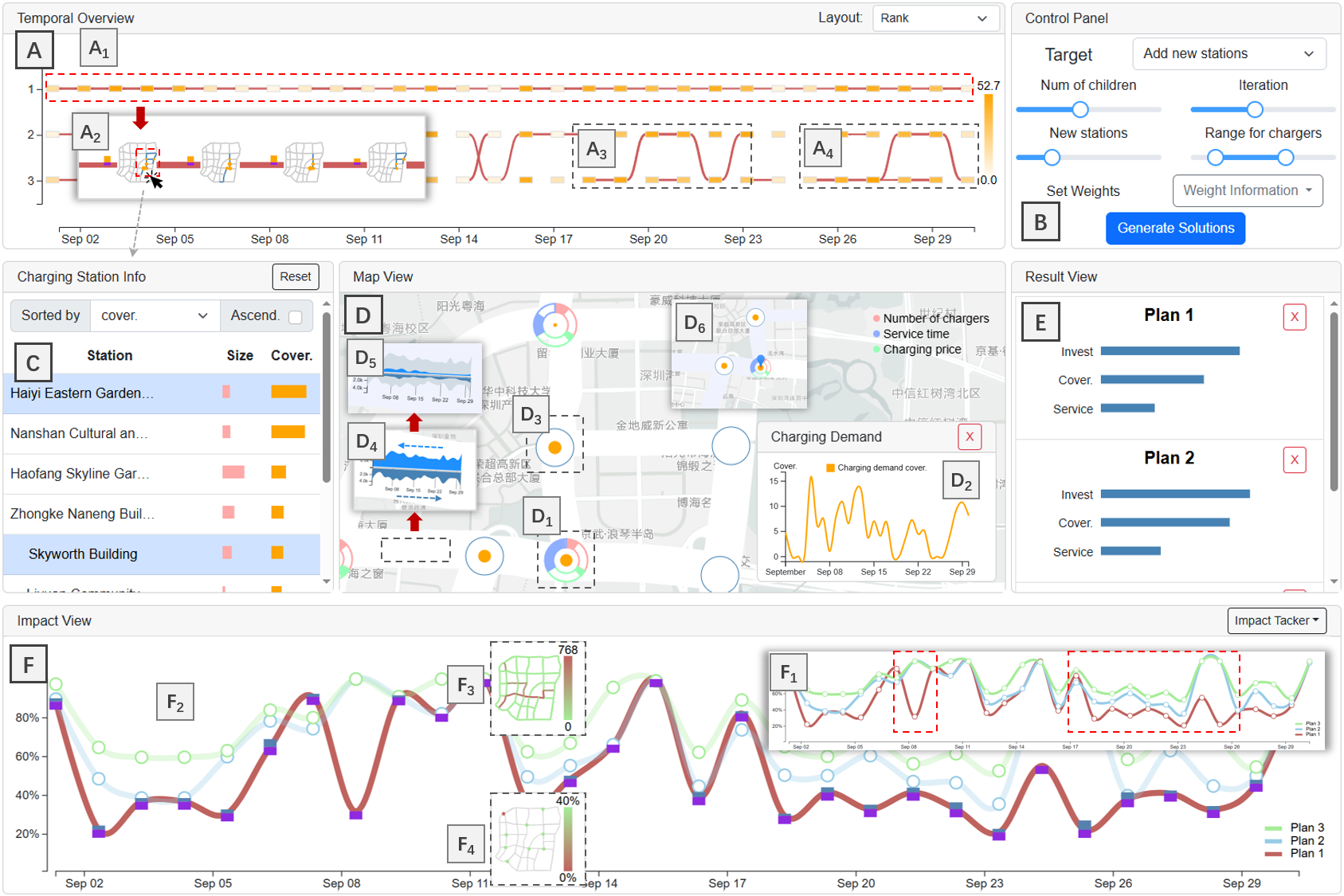}
  \vspace{-2.5mm}
  \caption{Our visualization system, named {\name}, facilitates the implementation of new charging stations within the interconnected transportation and power networks. The {\temporaloverview} (A) offers a condensed visual representation of fluctuations in traffic hotspots and charging demand. In the {\controlpanel} (B), users can adjust parameters to devise optimal strategies for deploying new charging stations. The {\chargingstationinfo} (C) module provides essential details about existing charging stations. The {\mapview} (D) furnishes detailed information on traffic volume, charging demand, and the operational status of charging stations. In the {\resultview} (E), users gain insights into the performance metrics of potential solutions. Lastly, the {\detailview} (F) enables users to compare various solutions, evaluating their respective impacts on both the road network and the power grid.}
  \label{fig:teaser}
  
  \vspace{-1mm}
}

\graphicspath{{figs/}{figures/}{pictures/}{images/}{./}} 

\usepackage{tabu}                      
\usepackage{booktabs}                  
\usepackage{lipsum}                    
\usepackage{mwe}                       
\usepackage{paralist}
\usepackage{balance}

\usepackage{mathptmx}                  

\begin{document}



\maketitle

\section{Introduction}
\par Over recent years, there has been a steady rise in the presence of electric vehicles (EVs) on our roads, with their expansion heavily reliant on the availability of charging infrastructure~\cite{unterluggauer2022electric,metais2022too}. However, ensuring the establishment and upkeep of a well-designed and resilient charging station network entails meticulous consideration of various factors. These factors encompass significant resource allocation~\cite{nicholas2019estimating}, effective coordination with pertinent transportation networks and power grids~\cite{shaukat2018survey}, among others. Despite these efforts, research highlights a prevalent issue: the underutilization of existing charging station infrastructure~\cite{bayram2023statistical,borlaug2023public}. Such inefficiencies not only lead to resource depletion but also exert a ripple effect on the interconnected transportation networks and power grids. Therefore, optimizing the deployment of charging station infrastructure calls for the urgent implementation of more sophisticated decision support systems. These systems aim to minimize investment costs while maximizing social benefits.

\par Researchers have conducted extensive studies in the field of charging station location problem (CSLP) to improve its reliability. Commonly employed methods include node-based and flow-based models~\cite{kchaou2021charging}. Node-based models assume that charging demand takes place directly at specific sites, while flow-based models assume that charging demand occurs while EVs are traveling. In recent years, there has been a growing focus on research concerning the coupled transportation and power networks in CSLP~\cite{unterluggauer2022electric,ahmad2022optimal}, which examines the conditions of both traffic and power grids. This integrated approach considers a broader spectrum of factors, which not only aids in the deployment of charging stations but also holds significant potential for reducing social costs and enhancing social welfare~\cite{unterluggauer2022electric}. However, despite numerous representative studies addressing CSLP~\cite{kchaou2021charging,unterluggauer2022electric,ahmad2022optimal}, they primarily focus on generating deployment results without analyzing spatiotemporal variations in transportation networks and power grids~\cite{zhou2021analyzing}, leading to incomplete evaluations of deployment results and inadequate support for decision-making rationality. Furthermore, existing charging station location models simplify computational complexity by making simplistic assumptions and considering limited factors~\cite{kchaou2021charging}, which may not adequately support expert decision-making in practical applications. Therefore, a method is required that seamlessly integrates multiple factors of concern for domain experts into the model deployment plan, while also elucidating the spatiotemporal changes it triggers. This approach aims to bolster the reliability and soundness of decisions regarding the deployment of charging stations.

\par Visual analytics techniques integrate intelligent recommendation models and human-centered multifaceted evaluations to address facility location problems (FLP)~\cite{deng2023survey}. Previous research has predominantly focused on traffic conditions as pivotal factors in FLP, including considerations such as traffic volume and speed for billboards~\cite{liu2016smartadp}, reachability for housing location~\cite{weng2018homefinder} and the scope of reachable destinations for warehouse siting~\cite{li2020warehouse}. However, there has been limited exploration of the implications of deployment results on the existing context. While some studies, like~\cite{chen2023fslens}, examined the effects on existing layouts following the siting of new fire stations, they merely redistributed historical fire incidents without considering broader network-level influences like traffic conditions. It's crucial to acknowledge that deploying new charging stations will affect both road networks and power grids. Charging stations can alter traffic volume by reshaping the travel routes of EVs~\cite{he2013optimal} and may strain power grids, potentially leading to overloads~\cite{yong2015review,unterluggauer2022electric}. Neglecting congestion and power grid strain can adversely affect the transportation and energy sectors. Therefore, comprehensive planning must balance economic concerns with social and environmental goals, assessing impacts on traffic flow, grid reliability, and community well-being. This holistic approach enables decision-makers to promote efficient and resilient infrastructure while minimizing negative societal effects. Consequently, CSLP necessitates analyzing both road networks and power grids, especially the post-deployment impact of new charging stations.

\par Nevertheless, the decision-making process for CSLP under coupled transportation and power networks presents several challenges: \textbf{1) Integrating Coupled Networks in the Same Context.} Despite the availability of numerous visualization solutions for road networks and power grids individually~\cite{deng2023survey,nga2012visualization,sanchez2018survey}, these solutions are primarily designed to address each network in isolation. However, to fully grasp coupled networks, it is imperative to analyze road networks and power grids concurrently. Separating the analysis of these networks could lead users to focus solely on one aspect, potentially overlooking the broader context. To the best of our knowledge, there remains a significant gap in visualization solutions directly tailored to scenarios involving coupled transportation and power networks. \textbf{2) Dynamic Spatiotemporal Patterns Evolution in Coupled Networks.} Previous research on CSLP has developed siting models considering factors from both road networks and power grids but has not illustrated the process of spatiotemporal changes in these networks~\cite{zhou2021analyzing}. This oversight results in an incomplete analysis and evaluation of deployment results, to some extent hindering the practical application of these models. The coupling of transportation and power networks implies many dependencies, where factors such as traffic generate charging demand and grid load influence the working status of charging stations. Effectively depicting these relationships is challenging yet essential for informed decision-making in CSLP. It enables planners to anticipate and mitigate potential challenges arising from dynamic spatiotemporal patterns, ensuring more robust and adaptive infrastructure planning strategies. \textbf{3) Post-Deployment Impact of New Charging Stations.} While some researchers have addressed the impact of placing new facilities~\cite{chen2023fslens}, their focus tends to be on resource allocation within specific areas, overlooking broader network-level effects. In our study, the addition of new charging stations affects both road networks and power grids, leading to issues such as vehicle congestion stemming from concentrated charging and increased grid load due to charging demand. Furthermore, the impact on coupled networks may dynamically evolve over time. Consequently, evaluating deployment strategies based solely on one or two criteria is inadequate. We advocate for the development of appropriate visualizations to effectively illustrate the post-deployment impact of new charging stations. Such features enable planners to make more informed decisions by considering the intricate interplay between transportation and energy systems.

\par In response to the aforementioned challenges, we introduce a visual analytics system, {\name}, designed to assess the existing configuration of charging stations and aid in the decision-making process for deploying additional stations. Our approach begins with an observational study, examining the methodologies currently utilized by domain experts for analyzing CSLP, while identifying their key concerns and expectations. Subsequently, we establish an integrated transportation and power network model to facilitate the evaluation of both the current charging station layout and the anticipated impact after deployment. Leveraging a genetic algorithm, we generate candidate solutions for charging station deployment. Furthermore, to enhance the representation of road network and power grid status, we employ a community discovery algorithm to identify traffic hotspots, employing various visual designs for effective communication. Finally, we conduct interviews with domain experts and execute two case studies to assess the efficacy of our system. Our contributions are delineated as follows:
\begin{compactitem}
    \item We have identified the analytical requirements of CSLP and introduced a methodology incorporating human-computer interaction and visualization.
    \item {\revised{We have refined several visualization designs to align with the analytical needs for CSLP, including the analysis of traffic hotspots and post-deployment impacts. Subsequently, we developed {\name}, a visual analytics system designed to guide users in making informed decisions about charging station deployment within the context of coupled networks.}}
    \item We have evaluated the feasibility and effectiveness of using the system {\name} for charging station deployment through two case studies and expert interviews.
\end{compactitem}

\section{Related Work}

\subsection{Charging Station Location Problem}
\vspace{-1mm}
\par The steady growth in the number of EVs relies on the availability of robust charging stations. Constructing charging stations demands substantial investment~\cite{nicholas2019estimating}, underscoring the need for comprehensive and reliable decision-making support.

\par In recent years, there has been a notable surge in global interest in CSLP. According to \cite{kchaou2021charging}, CSLP models can be categorized into two main approaches: \textit{node-based} and \textit{flow-based}. Node-based models~\cite{bouguerra2019determining,cui2019electric,dong2019electric,yi2022electric} employ the traditional FLP method, assuming that charging demand originates solely from nodes within the road network or specific areas within the city. These models are notably adept at facilitating planning for slow chargers. In contrast, flow-based models~\cite{guo2018battery,ghamami2020refueling,yang2022integrated} operate under the assumption that charging demand occurs during EV trips. Recently, there has been a growing emphasis among researchers on integrating transportation networks with power grids in urban planning~\cite{shaukat2018survey,mcgee2019state,unterluggauer2022electric}. Considering the coupling of the transportation network and the power network into CSLP not only expedites the deployment of charging stations and enhances decision-making rationality but also yields societal benefits at the urban system level~\cite{unterluggauer2022electric}. Consequently, recent studies on CSLP indicate a discernible trend towards the coupling of transportation and power networks~\cite{wang2018coordinated,lin2019multistage,unterluggauer2022electric,ahmad2022optimal}. \revised{In general, the factors to consider for CSLP, such as service time and integration with the power system~\cite{kchaou2021charging,lam2013electric,unterluggauer2022electric}, differ significantly from those involved in fossil fuel/service station deployment. These distinctions make CSLP a unique research domain.} 

\par Although researchers have made significant research progress in CSLP, \revised{current models still exhibit certain limitations.} First, charging station location models, particularly those considering coupled transportation and power networks, impose numerous constraints, resulting in \revised{computational complexity~\cite{kchaou2021charging}}. Second, these models often \revised{fail to include all real-world constraints} and researchers usually simplify their research scenarios~\cite{kchaou2021charging}. Furthermore, there is a prevailing focus on generating deployment results, with \revised{insufficient attention given to the spatiotemporal dynamics of traffic and electricity demand~\cite{zhou2021analyzing}}. These constraints highlight the inability of existing charging station location models to entirely supplant the decision-making expertise of domain experts. Additionally, there is a pressing need for more comprehensive evaluation methods, such as visual analytics, which can seamlessly integrate intelligent recommendation models with human-centered evaluations~\cite{deng2023survey}. Hence, unlike previous research, we explore CSLP through the lens of visual analytics. Introducing our novel framework, {\name}, we aim to bolster the reliability and efficacy of CSLP decision-making processes.

\subsection{Facility Location Problem based on Visual Analytics}
\par Currently, many researchers have adopted visualization techniques to address FLP from various perspectives. For instance, Liu et al.~\cite{liu2016smartadp} delved into billboard sitting, taking into account points of interest, traffic volume and vehicle speed. Weng et al.~\cite{weng2018homefinder} focused on accessibility-driven criteria when selecting sites for rental housing. In the context of warehousing, Li et al.~\cite{li2020warehouse} factored in logistics and transportation costs within specific regions, aiming to optimize routes to multiple destinations. Chen et al.~\cite{chen2023fslens} examined fire station placement, analyzing fire-related factors and simulating historical demand. Additionally, there are notable studies addressing store location problem~\cite{karamshuk2013geo,weng2018srvis} and ambulance station placement~\cite{li2015location}.

\par Typically, previous research has predominantly focused on devising solutions for FLP, often neglecting the potential consequences post-deployment of new facilities. To the best of our knowledge, \cite{chen2023fslens} stands as the sole endeavor to examine the potential impact following the establishment of new fire stations. However, their simulation merely replicated fire incidents in certain areas, lacking the capacity to assess the broader network-level effects. In our study, we address this gap by considering the coupling of transportation and power networks in CSLP. Collaborating with domain experts, we developed a coupled transportation and power network model. Introducing a novel visual analytics system alongside novel visual designs, we aim to scrutinize the status of the coupled network and evaluate the impact after the deployment of new charging stations.

\subsection{Geospatial Network Visualization}
\par The development of smart cities relies significantly on the intelligence of various networks, among which transportation networks and power grids stand as indispensable elements~\cite{amini2019distributed}. Currently, a plethora of visualization techniques are available for visualizing both transportation networks and power grids.

\par In the realm of transportation visualization, we organize relevant research based on several perspectives as outlined by~\cite{deng2023survey}. These perspectives include \textit{human mobility}~\cite{itoh2016visual,zeng2017visualizing,huang2019natural,yang2022epimob}, \textit{road network}~\cite{huang2015trajgraph,kamw2019urban}, and \textit{congestion analysis}~\cite{lee2019visual,wu2020towards,deng2021visual}. Concerning power grid visualization, Dao et al.~\cite{nga2012visualization} and Sanchez-Hidalgo et al.~\cite{sanchez2018survey} summarized commonly employed visualizations by smart grid researchers. These include parallel coordinates, matrix scatter diagrams, and Andrew's curves for analyzing multidimensional data within power grids. Furthermore, novel visualization designs have emerged, such as single-line diagrams with contouring~\cite{li2017distributed}, glyph-based visualization~\cite{gruchalla2023reevaluating}, and the epicentric cluster dendrogram~\cite{arunkumar2022pmu}.

\par To address the limitations observed in previous research, where the analysis often focuses solely on individual transportation or power grid networks without considering the overlay analysis of multiple networks, our study aims to introduce novel visual designs to simultaneously examine both road networks and power grids in urban environments. While previous studies have primarily focused on diverse transportation networks such as bus, tram, and metro systems~\cite{ding2018detecting,yildirimoglu2018identification,samant2021cowiz}, only a limited number have tackled multi-network analysis in urban scenarios.

\par For CSLP, it is crucial to simultaneously scrutinize temporal and spatial changes within both transportation and power networks. To achieve this goal, we propose a novel approach that integrates zoomable timeline visualization to provide an overview of traffic hotspots, charging demand, and grid load. Additionally, refined map and timeline visualization techniques are employed to illustrate the impact following the deployment of new charging stations. By employing these integrated visualization techniques, our study aims to provide more comprehensive decision-making support for the deployment of new charging stations.

\section{Observational Study}
\subsection{Experts' Current Practices and Expectations} \label{sec:experts-interview}
\par To gain a better understanding of the analytical requirements and experts' practices inherent in CSLP, we conducted interviews with five experts (\textbf{E1}-\textbf{E5}). These experts are from academia or industry and possess extensive experience in CSLP. \textbf{E1} and \textbf{E2} are from two different power infrastructure deployment companies, each accumulating over 8 years of practical experience in charging station deployment. Additionally, \textbf{E3}, an associate professor at a local university, contributes over a decade of research experience in the field of electric power systems. Furthermore, two researchers (\textbf{E4} and \textbf{E5}) bring two years of dedicated CSLP research experience, particularly focusing on modeling and problem-solving techniques. The discussions with experts have facilitated a deeper understanding of CSLP's current status and constraints, spanning both theoretical exploration and real-world applications. In light of these invaluable insights, we present a concise synthesis of the key findings, categorized into three primary dimensions:

\par \textbf{The Decision-making Process of Charging Station Deployment.} During our discussions, \textbf{E1} and \textbf{E2} elaborated on the decision-making process involved in charging station siting within real-world scenarios. Initially, they emphasized the importance of identifying candidate locations based on their experience. These potential sites are typically chosen for their proximity to high-traffic areas, which are crucial for attracting a steady flow of vehicles for charging purposes. Additionally, careful consideration is given to the condition of the power grid supply to ensure sufficient capacity for charging demand. Subsequently, they described the process of conducting thorough on-site inspections, taking into account various factors such as rental costs, power availability, and the overall environment surrounding the site. These meticulous assessments culminate in the final selection of the most suitable location for the charging station. Once the construction phase is completed, the focus shifts towards evaluating the feasibility of the siting scheme, primarily based on the operational performance of the charging station. It's worth noting that due to the substantial investment involved in constructing charging stations, relocation is seldom a viable option after the siting process. \textbf{E3} and \textbf{E5} provided additional insights regarding the implications of charging stations. In terms of traffic conditions, the presence of charging stations can influence vehicle flow and potentially impact surrounding traffic volume. Regarding power grids, two critical metrics, grid load and voltage, are indicative of the effects of charging stations. Grid load is directly influenced by charging demand. \textbf{E5} noted, ``\textit{When there's too much load or unbalance, it can mess with the voltage in the power grid, causing it to fluctuate or drop. And that could mess up how the charging stations normally work.}'' To sum up, the deployment of new charging stations involves various factors like traffic patterns and power grid infrastructure. Yet, it heavily depends on manual processes and practical experience, missing out on integrating data-driven methods to streamline decision-making and assess post-deployment impact effectively.

\par \textbf{Existing Tools for Addressing the Charging Station Location Challenge.} While there are several visualization systems available, their primary function is to monitor the operational performance and revenue generation of existing charging stations, rather than aiding in the siting process. ``\textit{We use navigation software to check out the traffic conditions at potential locations. And we've teamed up with a ride-hailing company before. They gave heat maps that help us find the busiest traffic spots in the city.}'' \textbf{E1} mentioned. \revised{Despite these existing tools, experts unanimously expressed the absence of dedicated assistance systems specifically tailored for charging station deployment, both within industry and academia.} ``\textit{In academia, researchers have come up with models and algorithms for deciding where to put charging stations. But here's the thing, they usually work with sample data just for their research.}'', said \textbf{E4}. As a result, there remains a gap between theoretical research and practical implementation within the field.

\par \textbf{Modeling the Coupled Transportation and Power Networks.} In the real world, both traffic and power grids are intricate systems with numerous factors at play, making it impractical to consider every detail. To effectively address the research problem and explore the interaction between traffic and the power grid, we consulted with experts to gain valuable insights into the coupled transportation and power networks. This coupling typically involves two key problem areas: traffic assignment problem and optimal power flow problem (which will be elaborated upon in \cref{sec:coupled-networks}). In terms of road networks, researchers often work with a sample road network or a simplified real urban road network. Furthermore, they are limited to a specific area of the city due to the computational complexity caused by constraints under consideration. As for power grids, their inherent complexity necessitates the use of simplified models for analysis. These may include widely used models such as the \textit{IEEE 14 Bus System}\footnote{\small{\url{https://labs.ece.uw.edu/pstca/pf14/pg_tca14bus.htm}}} and \textit{IEEE 33 Bus System}\cite{baran1989network}. In the context of CSLP, \revised{experts place greater emphasis on understanding the load and voltage of power grid nodes, with less focus on the intricate details of power grid structure or line specifications.} As \textbf{E3} noted, ``\textit{In real life, the layout of the power grid doesn't change much. So, if we're looking at a small study area, the differences in how the grid is set up won't really change our findings all that much.}''

\subsection{Experts' Needs and Expectations}
\par After gathering insights from interviews with the five experts (\textbf{E1}-\textbf{E5}), we delved into extensive discussions and iterations, resulting in the identification of the following key requirements.

\par \textbf{R.1 Provide a comprehensive overview of the current status of the road network and power grid.} Given the dynamic nature of transportation and power infrastructure, it's crucial to track their evolving status. By presenting key metrics over time, users can promptly detect issues such as grid overload and discrepancies between charging demand and grid capacity. Additionally, insights from interviews with \textbf{E1} and \textbf{E2} highlight a significant connection between traffic hotspots and charging demand. As \textbf{E1} remarked, ``\textit{It'd be helpful to give an overview to check out how traffic hotspots on the road network are changing.}''

\par \textbf{R.2 Investigate the detailed spatiotemporal variations in the road network and power grid.} While numerous algorithms for determining charging station locations have been proposed by researchers, these algorithms typically evaluate models from a broad perspective, such as investment, service, and charging demand coverage~\cite{kchaou2021charging}. While this approach allows for an evaluation of the pros and cons of charging station location plans, it fails to capture the specific changes in traffic flow, charging demand, and grid load over time and space, resulting in an insufficient assessment of the plan. \textbf{E3} noted, ``\textit{It is interesting to see the specific changes happening over time and space, and it's really valuable for digging deeper into the analysis,}''. Furthermore, experts expressed a desire to employ data-driven approaches to examine the proportional impact of road and power grid networks on potential deployment solutions, as well as the precise fluctuations in charging demand influenced by both networks.

\par \textbf{R.3 Analyze the impact of charging stations on both the road network and the power grid.} Based on insights from expert interviews and previous research, we have learned that charging stations can influence the traffic volume on surrounding roads by altering the route choices of EVs. Additionally, as noted by \textbf{E4}, charging stations can also affect the load on the power grid. Excessive or unbalanced grid load may affect the normal operation of charging stations due to voltage drops. To address this, a visualization scheme is needed to evaluate the impact of charging stations on both the power grid and road network. This will provide decision-makers with a comprehensive understanding of the problem-solving process within their models.

\par \textbf{R.4 Generate charging station siting solutions.} Given the current dependence on manual decision-making processes for siting new charging stations, there's an urgent need for a method to efficiently provide decision-makers with siting solutions. Furthermore, as highlighted by feedback from \textbf{E4} and \textbf{E5}, researchers commonly utilize convex optimization to solve CSLP, which yields a single solution constrained by numerous factors. However, what constitutes a desirable charging station location plan can vary greatly from person to person. Therefore, algorithms designed to generate siting solutions should aim to generate a diverse range of siting options.

\par \textbf{R.5 Conduct a comparative evaluation of various siting solutions.} Optimization algorithms designed for siting solutions typically consider multiple factors, leading to a range of optimization approaches with varying effects on the road network and the power grid. Consequently, decision-makers must undertake a thorough comparison and evaluation of different siting solutions. This analysis should determine whether the impact of these solutions on the coupled networks falls within acceptable bounds, enabling the selection of the most suitable solution.

\begin{figure*}[h]
  \centering
  \vspace{-3mm}
  \includegraphics[width=\textwidth]{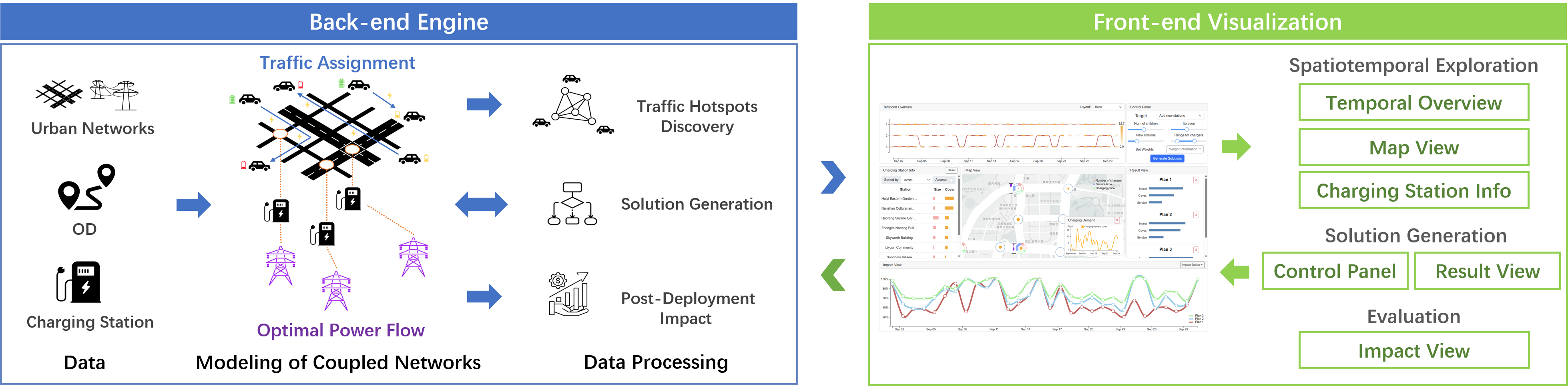}
  \vspace{-6mm}
  \caption{The system pipeline of {\name} includes both backend and frontend components. In the back-end engine, we formulate a model for the coupled transportation and power networks using three types of collected data. Users can generate candidate solutions for CSLP and analyze the post-deployment impacts. On the front-end visualization, we provide six coordinated views with interactive features. These views facilitate the exploration of current charging station layouts, the generation of deployment solutions, and the comprehensive evaluation of various alternatives.}
  \label{fig:pipeline}
\vspace{-4mm}
\end{figure*}

\section{{\name}}
\par Our system architecture, depicted in {\cref{fig:pipeline}}, comprises a back-end engine and a front-end visualization. The back-end engine inputs relevant data to solve the coupled transportation and power networks model. Leveraging techniques like community discovery and genetic algorithms, we conduct data analysis and generate solutions for charging station siting. On the front-end visualization, we have developed six components with full interactions, which enable users to efficiently analyze spatiotemporal data from the road network and the power grid, acquire deployment solutions, and evaluate these solutions more comprehensively. \revised{Experts have offered valuable insights throughout the iterative design process of our system, including data processing, system requirement identification, visual design, and evaluation.}

\vspace{-2mm}
\subsection{Back-end Engine}
\subsubsection{Data Description}
\vspace{-1mm}
\par Our system relies on three types of data: urban networks, origin-destination (OD) data, and charging station information. \textbf{1) Urban Networks.} To establish the transportation network, we sourced the primary road network data of a city from \textit{OpenStreetMap}\footnote{\small{\url{https://www.openstreetmap.org/}}}, including approximately 6,688 nodes (intersections) and 11,600 roads. For power network modeling, we employed the \textit{IEEE 14 Bus System}. \textbf{2) OD Data.} OD data captures trip origins, destinations, and associated trip volumes. Utilizing service data from ride-hailing and taxi services in the city, we acquired around 6 million OD data from September 1st to September 30th, 2019. \textbf{3) Charging Station Information.} We compiled data on over 1,000 charging stations within the city, including names, locations, and charger quantities (``\textit{Size}''). \revised{CSLP often sets intersections as candidate locations. To better integrate with the modeling of the coupled transportation and power networks, charging stations were geospatially aligned with the nearest intersections in the road network.}

\vspace{-2mm}
\subsubsection{Coupled Transportation and Power Networks} \label{sec:coupled-networks}
\par The modeling of coupled transportation and power networks involves two primary sub-problems: the traffic assignment problem (TAP) and the optimal power flow problem (OPF). \textbf{1) Traffic Assignment Problem.} Traffic assignment involves allocating OD pairs to a road network based on traveler's route choice criteria, \revised{which is widely used to predict traffic status~\cite{saw2015tap}. It operates on the principle of user equilibrium (UE)~\cite{glen1952some} to generate the traffic volume of each road in the road network}, where drivers seek to minimize travel time and cost during their trips. Charging demand arises as EVs traverse the network, accumulating at nodes. \textbf{2) Optimal Power Flow Problem.} The OPF seeks to determine the optimal operational status for power networks while adhering to power flow constraints or operational limits~\cite{frank2012optimal,abdi2017review}. In our study, we first identified the charging demand covered by existing charging stations. Subsequently, utilizing power grid parameters and charging demand, the OPF model calculates grid load, voltage and electricity price across various locations.

\par Inspired by \cite{wei2017network} and \cite{shao2022generalized}, we integrated the constraints of both the TAP and the OPF to formulate a coupled transportation and power network for CSLP. Employing the widely used commercial optimization solver \textit{Gurobi}\footnote{\small{\url{https://www.gurobi.com/}}}, we updated the status of both the road network and the power grid. Besides, to better identify traffic hotspots within the road networks, we employed the classic community discovery algorithm \textit{Louvain}~\cite{blondel2008fast} to partition the network into distinct areas based on the traffic volume of roads. Subsequently, we isolated the top-ranked areas exhibiting high traffic volume as traffic hotspots. Furthermore, we assessed the similarity between temporally adjacent traffic hotspots according to the number of sharing intersections. \revised{In terms of the power grid, the structure of power grids can be highly complex and irregular in reality. Some data, such as current, are indirect to assess the status of charging stations. Therefore, we focused on data directly related to the deployment of charging stations, specifically grid load and voltage, rather than displaying the entire power grid.}

\vspace{-2mm}
\subsubsection{Generate Solutions for Charging Station Deployment} \label{sec:genetic-algorithm}
\par To alleviate the workload of experts when selecting candidate locations, our system requires an algorithm capable of automatically generating solutions for CSLP. After careful consideration, we chose to employ a genetic algorithm due to its ability to ensure diversity among alternative solutions\cite{katoch2021review}. Candidate locations are designated as nodes within the road network lacking charging stations. Through a comprehensive literature review and expert interviews, we have incorporated multiple metrics into the objective function. These metrics include: \textbf{1) Charging demand coverage.} A location with high charging demand is often considered an optimal choice for establishing a new charging station. Charging demand at each location is determined by the coupled networks model detailed in \cref{sec:coupled-networks}, \revised{with charging demand coverage indicating the proportion of charging demand at a specific location relative to the total charging demand ($C_{cover}$)}. \textbf{2) Service time.} Minimizing the service time for EVs can significantly enhance service efficiency. As shown in \cref{equa:service},
\vspace{-4mm}
\begin{equation}
    \revised{C_{service}(x_i) = \frac{CD_i}{x_i} \label{equa:service},}
    \vspace{-2mm}
\end{equation}
to simplify the problem, we calculate the service time at location $i$ by dividing the charging demand $CD_i$ by \revised{the number of chargers ($x_i$)}. \textbf{3) Investment.} \revised{The more chargers at a new charging station, the higher the charging station investment ($C_{invest}$).} While deploying numerous chargers may reduce customer waiting times, it could lead to underutilization of individual chargers, resulting in wasted investment.

\par By incorporating these three metrics, the objective function is formulated as shown in \cref{equa:obj}:
\begin{equation}
    \revised{\min_{x{\in}\textbf{X}}{f_0(x)} = -\sum_i{\omega_{1}C_{cover}(x_i) + \omega_{2}C_{service}(x_i) + \omega_{3}C_{invest}(x_i)} \label{equa:obj}.}
    \vspace{-2mm}
\end{equation}
Recognizing that users may prioritize different optimization goals, {\name} affords users the flexibility to adjust the weights of each metric ($\omega_{1}$, $\omega_{2}$, and $\omega_{3}$) within the objective function to suit their preferences.

\par We integrated the coupled transportation and power networks with the genetic algorithm. Initially, we executed the coupled networks model to generate charging demand. Subsequently, we configured the parameters of the genetic algorithm, including constraints such as the number range of chargers, the number of new charging stations, the weight of metrics, and the number of iterations. The genetic algorithm then produced multiple candidate solutions. Finally, we re-executed the coupled networks model to assess the effects of the newly deployed solutions on both the road network and power grid.

\subsection{Front-end Visualization}
\par {\name} consists of six primary components. This section presents the visual design of each view within our system, alongside alternative designs that were under consideration. Adhering to the visualization mantra ``\textit{overview first, zoom and filter, then details-on-demand}''~\cite{shneiderman2003eyes}.

\begin{figure}[h]
  \centering
    \vspace{-3mm}
  \includegraphics[width=0.5\textwidth]{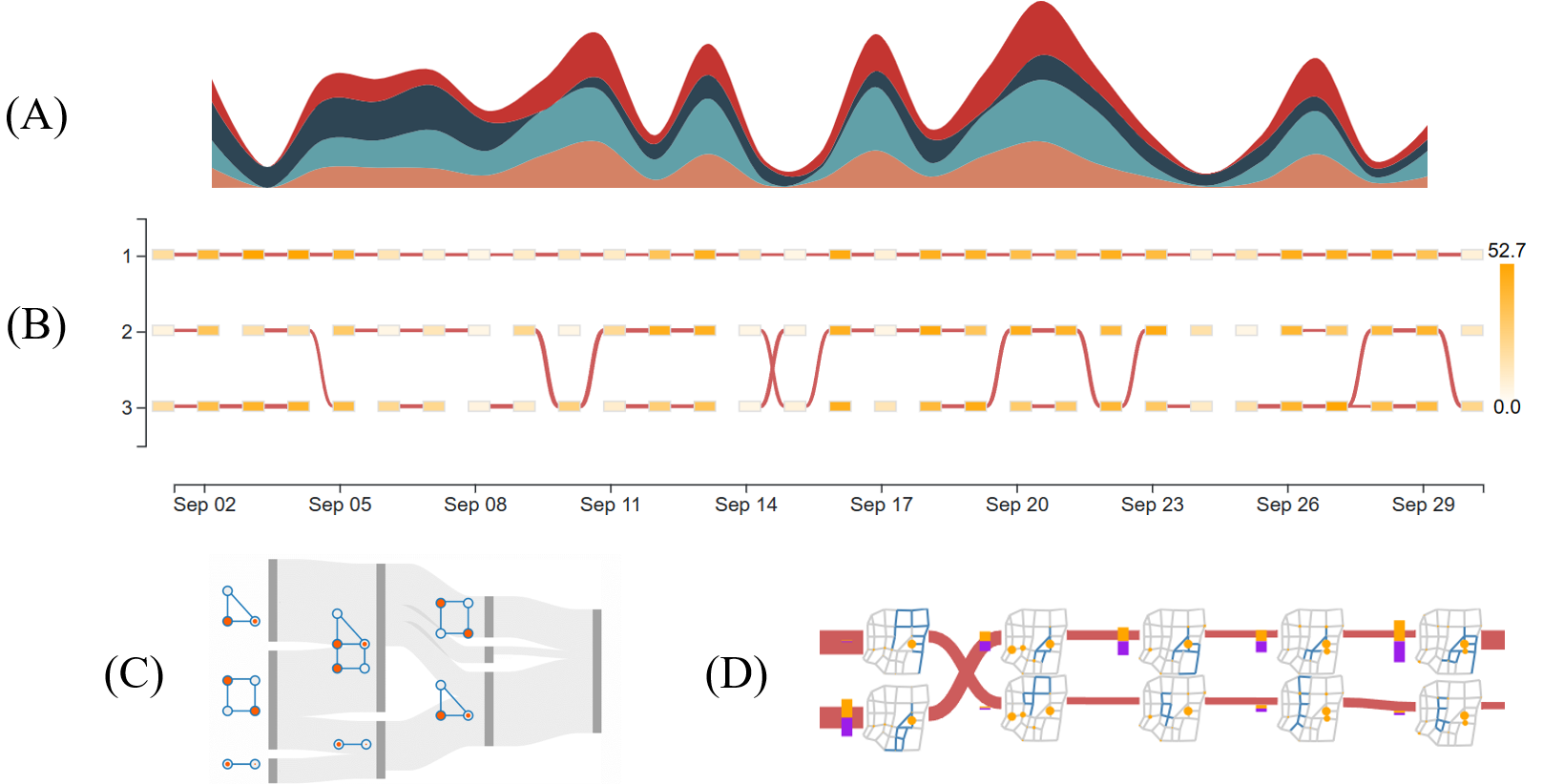}
  \vspace{-4mm}
  \caption{Design alternatives for the {\temporaloverview}. (A) \revised{Use a stacked area chart to show the charging demand across various traffic hotspots.} (B) Present the basic layout of the current {\temporaloverview}. (C) Showcase a glyph design depicting demand and traffic hotspots. (D) Demonstrate a glyph design featuring the road network, traffic hotspots, charging demand, and grid load.}
  \label{fig:DA-temproal}
    \vspace{-6mm}
\end{figure}

\vspace{-2mm}
\subsubsection{\temporaloverview}
\par The {\temporaloverview} in {\cref{fig:teaser}A} is designed to offer an overview of both the road network and the power grid (\textbf{R.1}), while illustrating the temporal evolution of traffic hotspots, charging demand, and grid load (\textbf{R.2}). As shown in {\cref{fig:teaser}$\textrm A_1$}, the x-axis encodes time and the y-axis encodes the rank of each traffic hotspot at different time points according to average traffic volume. Users can toggle between different y-axis values, such as the average traffic volume or the size of the traffic hotspot area, by selecting different layouts in the drop-down list at the top-right corner. Moreover, we offer the ``\textit{Link}'' layout to better demonstrate the connection among different traffic hotspots. Orange rectangles represent traffic hotspots, with varying opacity indicating the corresponding charging demand. The width of the red lines indicates the similarity between two traffic hotspots. Upon zooming in, as shown in {\cref{fig:teaser}$\textrm A_2$}, each orange rectangle transforms into a glyph to offer more detail. Each glyph comprises a road network and two bars on the left. Blue roads in the network represent the roads within the respective traffic hotspots, while the radius of orange circles reflects the charging demand at each location. The height of the orange bars represents the charging demand coverage within the traffic hotspots relative to the total charging demand. Similarly, the height of the purple bars encodes the average grid load within the traffic hotspot, representing the proportion of charging demand served by the charging stations in the hotspot compared to the total charging demand. If the orange bar in a glyph is higher than the purple bar, it suggests some charging demand is not served by the charging stations in the traffic hotspot.

\par \textbf{Design Alternatives.} {\cref{fig:DA-temproal}} presents several alternative designs that were evaluated before adopting the current design. \revised{In {\cref{fig:DA-temproal}A}, a design utilizing stacked area charts to represent traffic hotspots is depicted.} Different colors represent different traffic hotspots, while the width of the bars indicates the charging demand. However, this approach encounters challenges in consistently assigning colors due to the varying areas of traffic hotspots over time. Additionally, it fails to effectively illustrate the connections between different traffic hotspots. Consequently, we opted for the layout depicted in {\cref{fig:DA-temproal}B} as the basic design for the {\temporaloverview}. Regarding glyph design, we first modified the design proposed by {\cite{vehlow2015visualizing}} in {\cref{fig:DA-temproal}C}. While this design offers a direct representation of traffic hotspot structure, it cannot precisely depict the position of traffic hotspots within the road network. Conversely, the design presented in {\cref{fig:DA-temproal}D} incorporates the road network as the background, addressing this limitation. Additionally, two bars are introduced on the left side of the glyph to provide detailed information about charging demand and grid load.

\vspace{-2mm}
\subsubsection{\chargingstationinfo}
\par {\Cref{fig:teaser}C} complements the fundamental overview of the current charging stations (\textbf{R.1}), providing experts with insights into the service status of existing charging stations. We present a tabular format in {\cref{fig:teaser}C}, showcasing two types of information concerning the charging stations in the area: the number of chargers (``\textit{Size}'') and the average coverage (rate) of charging demand (``\textit{Cover.}''). Bar charts are employed to visually represent both the size and charging demand, with pink bars indicating the size of the charging station and orange bars indicating the coverage rate of charging demand. Within the {\chargingstationinfo} section, users have interactive functionalities enabling them to sort charging stations based on size or charging demand coverage. \zyt{When users hover on a bar, the data value of the bar will show up.} Additionally, they can choose to rank the data in ascending order by utilizing the ``\textit{Ascend.}'' checkbox. Furthermore, \revised{when users select a traffic hotspot in the {\temporaloverview}, the charging stations in the corresponding traffic hotspot will be highlighted.} Upon selecting a charging station of interest from the list, the {\mapview} automatically navigates to the corresponding charging station.

\vspace{-2mm}
\subsubsection{\mapview}
\par The {\mapview} ({\cref{fig:teaser}D}) is designed to provide comprehensive insights into each charging station, including spatiotemporal patterns in traffic conditions, charging demand and information about charging stations (\textbf{R.1} and \textbf{R.2}), as well as the impact of charging stations on the road network and power grids (\textbf{R.3}). This component comprises two primary sections: \textit{charging stations} and \textit{road map}.

\par \textbf{Charging Stations.} As shown in {\cref{fig:teaser}$\textrm D_1$}, a glyph is used to convey information about charging stations. \revised{The circular design is suitable to locate the glyph at each intersection.} The radius of the inner orange circle indicates the average charging demand of the charging station. Divided into three sectors, it conveys three types of information: \textit{size}, \textit{electricity price}, and \textit{service time}, with larger sector areas indicating larger corresponding values. To demonstrate the operational status of charging stations, a purple violin plot illustrates the voltage distribution at this location on the left side of the glyph, with two black lines denoting the upper and lower limits of the voltage. The purple area's width shows the approximate voltage frequency across different data value ranges. \revised{We do not adopt an embedded design to reduce the complexity of the glyph.} To provide detailed charging demand of each station, when users select a charging station on the map or within the {\chargingstationinfo}, a pop-up window will appear displaying a line chart in {\cref{fig:teaser}$\textrm D_2$}. The x-axis encodes time while the y-axis encodes the coverage of charging demand for the charging station.

\par \textbf{Road Map.} Apart from using the glyph to illustrate information about charging stations, we employ other visualizations to depict the charging demand at each location and the traffic volume on each road. Similarly, the radius of the orange circle represents the average charging demand of the location (\cref{fig:teaser}$\textrm D_3$). We use area charts with blue hues to demonstrate traffic volume. Upon zooming in, these area charts are centered on each road (\cref{fig:teaser}$\textrm D_4$). The x-axis of the area chart indicates time, while the y-axis represents traffic volume. Distinct shades of blue are chosen to signify traffic volumes in different directions of the road. To aid users in quickly assessing the connection between a charging station and its surrounding traffic condition, clicking on a charging station triggers the highlighting of the associated roads (\cref{fig:teaser}$\textrm D_6$), accompanied by the transformation of area charts into {\cref{fig:teaser}$\textrm D_5$}. Area charts in light gray represent the total traffic volume in one direction, while those in clear blue denote the vehicles that may charge at the selected charging station.

\vspace{-2mm}
\subsubsection{{\controlpanel} and {\resultview}}
\par The {\controlpanel} ({\cref{fig:teaser}B}) and {\resultview} ({\cref{fig:teaser}E}) are designed to facilitate experts in generating and assessing solutions of CSLP (\textbf{R.4} and \textbf{R.5}). Within the {\controlpanel}, users initially specify the deployment target, such as adding new charging stations or adjusting charger quantities. Subsequently, users can fine-tune various parameters related to the genetic algorithm, including the number of children per iteration, iteration count, new charging station quantity, and charger quantity range at each location. Increasing the number of children per iteration and iterations enhances solution diversity and optimality, albeit at the expense of increased runtime. Additionally, users have the flexibility to adjust the weights of three metrics in the objective function (\cref{sec:genetic-algorithm}). After clicking the ``\textit{Weight Information}'' button, users can use drop-down sliders to adjust the weight of different metrics.

\par Upon activating the ``\textit{Generate Solutions}'' button in {\cref{fig:teaser}B}, the {\controlpanel} initiates the generation of candidate solutions. \revised{Considering the position of the module and inspired by \cite{chen2023fslens}, we adopt a list to display different solutions.} In the {\resultview}, the top three candidate solutions are presented based on the objective function. Each solution is accompanied by a bar chart depicting three crucial metrics: \textit{investment}, \textit{charging demand coverage}, and \textit{service time}. Further details of these three solutions are provided in the {\detailview}. After users select a solution, the data in the {\mapview} and the {\detailview} will be updated. As shown in {\cref{fig:case-2}$\textrm C_2$}, new charging stations will be highlighted with red markers in the {\mapview}.

\vspace{-2mm}
\subsubsection{\detailview}
\par The {\detailview}, depicted in {\cref{fig:teaser}F}, serves the purpose of providing experts with details of the schemes generated by the system, enabling evaluation of solutions for new charging station deployment (\textbf{R.5}) and assessing the post-deployment impact (\textbf{R.3}). In the context of CSLP, charging demand stands out as a pivotal metric to consider. Therefore, the line chart in {\cref{fig:teaser}$\textrm F_1$} illustrates the temporal evolution of charging demand coverage across various solutions, with time points marked by circles. As users seek a solution with optimal charging demand, they also need to explore the potential impact on the road network and power grid to conduct a thorough evaluation.

\par In {\cref{fig:teaser}$\textrm F_2$}, upon selecting a target solution, the corresponding curve is highlighted, and the circles representing time points transform into dual bars. The blue bar signifies the impact on the road network, while the purple one denotes the impact on the power grid. The height of these bars reflects the extent of influence (e.g., how many roads/nodes of the power grid are affected), indicating the magnitude of impacts. For a detailed analysis of the specific effects on both the road network and the power grid, users can click on the bars, prompting the system to display the spatial distribution of impact on the road network ({\cref{fig:teaser}$\textrm F_3$}) and the power grid ({\cref{fig:teaser}$\textrm F_4$}) at the corresponding time point. The color of roads encodes the change in traffic volume while the color of nodes represents the change in voltage. In addition, users can use the ``\textit{Impact Tracker}'' at the top-right corner of the {\detailview} to filter the range of impacts they are interested in.

\par
\begin{figure}[h]
  \centering
    \vspace{-3mm}
  \includegraphics[width=0.5\textwidth]{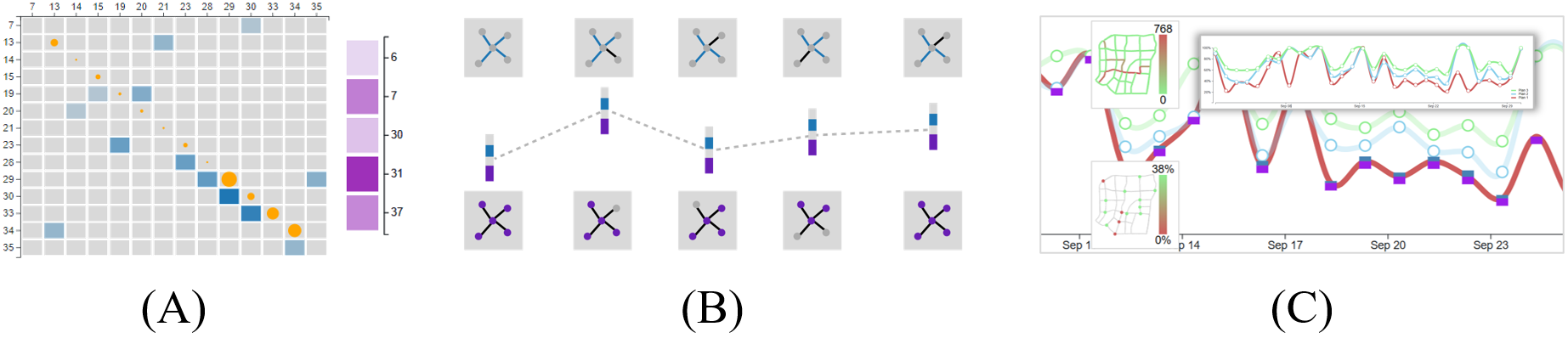}
  \vspace{-6mm}
  \caption{Design alternatives of the {\detailview}. (A) An adjacency matrix method showcasing traffic volume, charging demand and voltage. (B) A line chart integrated with road maps illustrating the impact after deployment. (C) A combined line chart and road map visualization illustrating both the impact after deployment and charging demand simultaneously.}
  \label{fig:DA-detailview}
    \vspace{-3mm}
\end{figure}

 \textbf{Design Alternatives.} Prior to adopting the current design, we explored several alternative approaches. Initially, we devised a layout based on an adjacency matrix, as illustrated in {\cref{fig:DA-detailview}A}. The left matrix represents the adjacency matrix of the road network, while the orange circles indicate charging demand at each location. The purple squares on the right encode voltage. However, this approach proved inefficient due to the sparsity of connections within road networks, leading to wasted space. Additionally, the matrix lacked intuitive road identification. Subsequently, recognizing the temporal variability of impacts, we experimented with the design shown in {\cref{fig:DA-detailview}B}. Here, two rows of road maps depicted the impact of new charging stations on the road network and the power grid, utilizing bars and a gray line chart to denote impact magnitude. Yet, redundancy between the line chart and bars, along with difficulties in comparing charging demand across solutions, prompted us to revise the line chart to represent charging demand per solution.

\section{Evaluation}
\subsection{Case Studies} \label{sec:case-study}
\xlw{We invited experts \textbf{E1}-\textbf{E5} (mentioned in \cref{sec:experts-interview}) to assess the effectiveness and usability of the system. We further collected their feedback for future improvements. The following two case studies demonstrate the activities performed by \textbf{E1}-\textbf{E5}.}

\begin{figure*}[h]
  \centering
    \vspace{-3mm}
  \includegraphics[width=\textwidth]{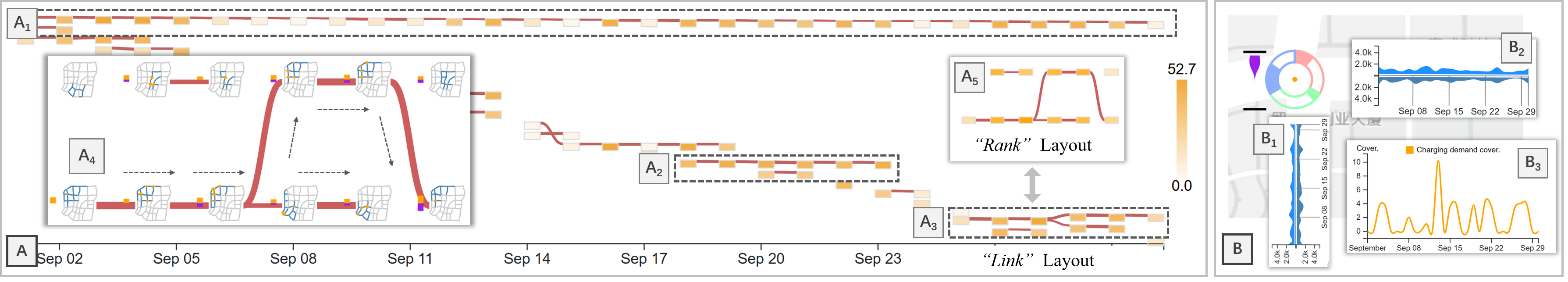}
  \vspace{-6mm}
  \caption{Experts’ operations in case I. (A) Experts used the ``\textit{Link}'' layout and the ``\textit{Rank}'' layout to illustrate the connection among different traffic hotspots. (B) The surrounding traffic condition of a charging station with low charging demand.}
  \label{fig:case-1}
    \vspace{-3mm}
\end{figure*}

\subsubsection{\textit{Case I: Evaluate the Current Layout of Charging Stations.}}
\par To explore the current layout of charging stations, experts selected an economic center within the city as the target area, characterized by numerous high-tech enterprises and a local university. 

\par \textbf{Explore the Overview of the Target Area.} First, experts focused on the spatiotemporal data changes in the {\temporaloverview}. As shown in {\cref{fig:teaser}$ \textrm A_1$}, the opacity of each orange rectangle indicates the charging demand of the corresponding traffic hotspot. The rectangles with high opacity show the corresponding traffic hotspot has high charging demand. Generally, the charging demand fluctuates with time. \textbf{E3} selected the ``\textit{Link}'' layout to better observe the connection among different traffic hotspots. According to the red lines, {\cref{fig:case-1}$\textrm A_1$} illustrates some traffic hotspots that have been connected for an entire month, which shows a strong connection. Many orange rectangles on this connection also have high opacity, indicating that the charging demand of these traffic hotspots is relatively high. {\cref{fig:case-1}$\textrm A_2$} and {\cref{fig:case-1}$\textrm A_3$} revealed two groups of representative traffic hotspots emerging from the middle and the end of this month, also characterized by tight connectivity and high charging demand. \textbf{E3} remarked, ``\textit{It seems that high charging demand has a certain correlation with these traffic hotspots.}'' Subsequently, some experts employed the ``\textit{Rank}'' layout in {\cref{fig:teaser}A}, where the y-axis represents the rank of traffic volume in different traffic hotspots.  They found that the traffic hotspots highlighted by the upper red rectangle in {\cref{fig:teaser}$\textrm A_1$} not only have been connected throughout the month but also the traffic volume consistently ranked first. \textbf{E2} commented, ``\textit{The traffic volume in these areas has been consistently high throughout this month and often become a traffic hotspot. This suggests further exploration.}'' 

\par After switching between two layouts of the {\temporaloverview}, the experts wanted to check the traffic hotspots in detail. As shown in {\cref{fig:teaser}$\textrm A_2$}, the orange rectangles turned into the glyph after zooming in. {\cref{fig:teaser}$\textrm A_2$} demonstrates some representative examples of the traffic hotspots ranking first in {\cref{fig:teaser}$\textrm A_1$}. According to the highlighted blue roads, these traffic hotspots are located on the right of the road network. The bars on the left of the glyph encode the charging demand (the orange bar) and the grid load (the purple bar) of these traffic hotspots. However, the orange bars are higher than the purple bars in {\cref{fig:teaser}$\textrm A_2$}, which indicates although charging demand is high, the actual demand served by charging stations is small. Similarly, {\cref{fig:case-1}$\textrm A_4$} illustrates another group of traffic hotspots in the top-left of the road network, which refers to {\cref{fig:case-1}$\textrm A_3$} and {\cref{fig:case-1}$\textrm A_5$} in two different layouts.  According to the arrows in {\cref{fig:case-1}$\textrm A_4$}, these traffic hotspots start with an area in the top-left of the road network and then split into two different traffic hotspots. Likewise, the long orange bars and many orange circles indicate these traffic hotspots have high charging demand, but the purple bars on the left of the glyph are still low, which means existing charging stations don't fully leverage the current charging demand.

\par \revised{Besides reviewing the {\temporaloverview}, some experts also utilized the {\chargingstationinfo} to examine the size and the charging demand of existing charging stations.} They sorted charging stations by charging demand coverage in {\cref{fig:teaser}C}. According to the length of the orange bars, some charging stations have high charging demand, while others are relatively low. \textbf{E2} checked the lengths of the red bar and pointed out, ``\textit{Some charging stations are large in scale, but their charging demand may not be correspondingly high, indicating potential resource wastage.}'' This is also another reason why existing charging stations may not adequately meet the current charging demand.

\par \textbf{Analyze Detailed Information of the Existing Charging Stations.} Continuing, experts moved to the {\mapview} to examine the map. They moved to the traffic hotspot suggested by {\cref{fig:teaser}$\textrm A_2$}. Firstly, they looked at the charging demand of different locations in the road network. Based on the radius of the orange circles, there were some locations with substantial charging demand but no charging stations ({\cref{fig:teaser}$\textrm D_3$}). \textbf{E2} discovered a charging station with high charging demand in {\cref{fig:teaser}$\textrm D_1$}, indicated by a large orange circle. The size of the charging station is relatively small according to the red sector, however, ``\textit{The area of the blue sector is large, it seems this charging station has significant service pressure.}'', said \textbf{E2}. The purple violin plot displays the distribution of the voltage of this charging station. Due to the high service pressure, sometimes the voltage of this charging station can be exceptionally low, even close to the lower limit. In contrast, nearby in the same region, the charging station in {\cref{fig:case-1}B} has low charging demand according to the small orange circle and the line chart in {\cref{fig:case-1}$\textrm B_3$}. \textbf{E2} remarked,``\textit{This indicates that the deployment of charging stations in this area is not rational enough, which may result in load unbalance to the power grid. If we can deploy some new charging stations properly, it would be beneficial for improving the load balance.}''


\par Subsequently, experts continued to examine the surrounding traffic conditions of charging stations. After zooming in the {\mapview}, the area charts displaying traffic volume for each road were revealed. The area chart in {\cref{fig:teaser}$\textrm D_4$} shows the traffic volume on roads adjacent to the charging station with high charging demand. \textbf{E3} then clicked on this charging station and the roads related to this charging station were highlighted in {\cref{fig:teaser}$\textrm D_6$}. The area charts also turned into {\cref{fig:teaser}$\textrm D_5$}, where the total traffic volume area turns gray and the blue area represents the vehicles that may charge at this charging station. Conversely, for charging stations with relatively low charging demand in {\cref{fig:case-1}B}, the area charts demonstrating the traffic volume are also low ({\cref{fig:case-1}$\textrm B_1$ and $\textrm B_2$}). \textbf{E3} remarked, ``\textit{There appears to be a positive correlation between charging demand and traffic volume.}''

\par To sum up, after the operations in case I, experts agreed that: (1) Some traffic hotspots have high charging demand. (2) The layout of the current charging stations is not reasonable enough. Some charging stations are overloaded, while some are idle, which may cause load unbalance in the power grid. (3) There is still much potential charging demand in this area. Properly locating new charging stations may utilize the charging demand and balance the load on the power grid.

\begin{figure*}[h]
  \centering
    \vspace{-3mm}
  \includegraphics[width=\textwidth]{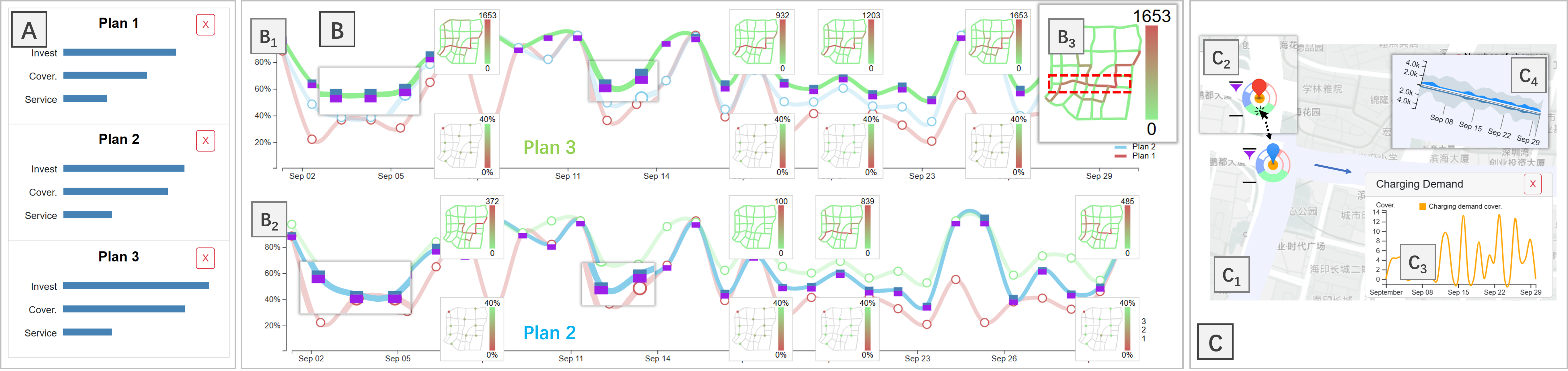}
  \vspace{-6mm}
  \caption{Experts' operations in case II: (A) The overall information of each candidate solution for charging station deployment. (B) Compare and evaluate the impact after the deployment of new charging stations. (C) An example to explore the location of a new charging station.}
  \label{fig:case-2}
    \vspace{-3mm}
\end{figure*}

\subsubsection{\textit{Case II: Deploy New Charging Stations.}}
\par \textbf{Generate Solutions for Charging Station Deployment.} Based on the findings in case I, experts considered deploying new charging stations in this area. \textbf{E1} and \textbf{E2} set parameters in the {\controlpanel}. Each solution will add two new charging stations to the road network and the chargers in each charging station are limited in [6,20]. After clicking the ``\textit{Generate Solutions}'' button, some candidate solutions are listed in {\cref{fig:case-2}A}. Each solution is more optimal in different metrics. For example, Plan 1 emphasized achieving relatively shorter service time and less investment. Plan 2 is a relatively balanced solution, demonstrating decent performance across all metrics. Plan 3 has the highest charging demand coverage, but its investment is also the highest.

\par \textbf{Evaluate and Compare Different Solutions.} The {\detailview} demonstrates the changes in charging demand coverage of each solution over time. As highlighted by red rectangles in {\cref{fig:teaser}$\textrm F_1$}, Plan 1 shows significant gaps compared to the other two plans. ``\textit{The deployment of charging stations needs to pursue a return on investment. Although Plan 1 has a smaller investment, it covers the least charging demand.}'', said \textbf{E1}. Since the trends of Plan 2 and Plan 3 are similar, \textbf{E4} wanted to further evaluate the impact after deployment. Because ``\textit{The increase in traffic volume can easily lead to congestion}'', so he first used the drop-down sliders in the ``\textit{Impact Tracker}'' to filter the roads where the traffic volume increased after deployment. He then clicked on the line charts to check the impact on the road network and the power grid. {\cref{fig:case-2}$\textrm B_1$} illustrates the impact of Plan 3. The heights of the blue bar on the line chart change with time, which indicates the impacts on the road network are different. \textbf{E4} selected several time points when more significant changes occurred in the road network. The glyph on the top of {\cref{fig:case-2}$\textrm B_1$} demonstrates the road where traffic volume has increased in Plan 3. Some roads in the middle of the road network are marked in red, indicating the traffic volume of these roads increases significantly.  \textbf{E1} explained, ``\textit{Some of these roads, especially the roads running in the east-west direction (highlighted by the red rectangle in {\cref{fig:case-2}$\textrm B_3$}), are part of an arterial road. This arterial road serves as one of the main commuting routes in the city and is usually congested. It appears that Plan 3 may lead to an increase in traffic pressure on the arterial road, resulting in further congestion.}'' The impact of Plan 2 is demonstrated in {\cref{fig:case-2}$\textrm B_2$}. Generally, the heights of the blue bars on the line chart of Plan 2 are lower than Plan 3, which means Plan 2 has less impact on the traffic condition. As shown in the road network on the top of {\cref{fig:case-2}$\textrm B_2$}, Plan 2 affects more scattered roads. Although sometimes Plan 2 also increases the traffic volume of the congested arterial road, it only affects some parts of it. Moreover, the increase in traffic volume is much less than Plan 3. As for the impact on the power grid, the heights of the purple bars on the line charts in {\cref{fig:case-2}$\textrm B_1$} and {\cref{fig:case-2}$\textrm B_2$} are similar. The glyph at the bottom of {\cref{fig:case-2}$\textrm B_1$} and {\cref{fig:case-2}$\textrm B_2$} suggests the voltage of some locations has increased, ranging from 15\% to 40\%. \textbf{E4} said, ``\textit{This could be because the new charging stations are beneficial for balancing the load on the power grid. A more balanced load distribution would lead to a more stable power grid.}''.

\par \textbf{Explore New Charging Stations in the Map.} After comparing different solutions, \textbf{E3} moved to the {\mapview} to check the locations of new charging stations in Plan 2. For example, {\cref{fig:case-2}C} shows the location of a new charging station close to the arterial road in {\cref{fig:case-2}$\textrm B_3$}, where the blue arrow in {\cref{fig:case-2}$\textrm C_1$} encodes its direction. The large orange circle indicates this charging station has high charging demand. The purple violin plot shows the voltage of this charging station remains close to the upper limit, which indicates this charging station operates normally. When \textbf{E3} clicked on the charging station, the roads related to the charging station were highlighted by light blue in {\cref{fig:case-2}$\textrm C_1$}. Afterward, \textbf{E3} zoomed in to examine the traffic volume of the arterial road. In {\cref{fig:case-2}$\textrm C_4$}, the area charts in light gray indicate the total traffic volume of the arterial road, while other area charts in blue represent the traffic flow related to this new charging station. This indicates around 5\%-10\% of the vehicles on the arterial road may charge at this charging station. \textbf{E3} remarked, ``\textit{This function can be useful to analyze the attraction of a new charging station.}''

\par In summary, experts used {\name} to generate and evaluate different solutions for new charging station deployment. They believed that Plan 2 is relatively ideal because it can meet more charging demand, balance the load on the power grid, and may not significantly increase the traffic pressure on an important arterial road.

\subsection{Interview with Domain Experts}
\par \textbf{Procedure.} To validate the effectiveness and usability of our system, we conducted interviews with \textbf{E1}-\textbf{E5}, as introduced in \cref{sec:experts-interview}. First, we briefly introduced {\name} and provided a simple tutorial to demonstrate the visual design and interaction. Later on, experts could explore the system for about half an hour. \revised{Given that experts are unfamiliar with visual analytics, we first adapted the co-discovery method to help them learn how to use {\name}. Additionally, we provided other laptops to allow experts to use the system individually.} Some of their operations are shown in \cref{sec:case-study}. Subsequently, we conducted an individual interview for each expert. Each interview lasted approximately 20 minutes and we collected their feedback and suggestions.

\par \textbf{System Design.} Overall, experts approved that {\name} is a useful tool to facilitate the deployment of new charging stations. Following the simple tutorial, they had no trouble comprehending the purpose and meaning of our visual design. \textbf{E1} and \textbf{E2} expressed that in the industry, traditional methods for selecting charging station locations require significant manpower and experience to investigate and select candidate locations. \textbf{E2} mentioned, ``\textit{The criteria for determining why a location is suitable can be ambiguous, and sometimes even experience can be misleading.''} With {\name}, he can rapidly get an overview of the target area and generate candidate solutions. ``\textit{The system provides the traffic volume around a charging station, as well as the change in traffic hotspots and charging demand, which are valuable for decision-making.}'', he added. \textbf{E3}-\textbf{E5} also expressed their affirmation of our system. \revised{\textbf{E5} noted, ``\textit{The {\detailview} displayed both the possible positive and negative impact after the deployment.}''} In particular, they were interested in the {\temporaloverview}. After our explanations, they found it not only provided different analysis alternatives, such as the ``\textit{Link}'' layout and other layouts based on traffic conditions, but was also capable of providing them with a quick overview of the road network and the power grid. \textbf{E3} said, ``\textit{It indeed demonstrates the spatiotemporal changes in traffic hotspots and charging demand.}'' Besides, following the interview, he noted that our system provided them with many new insights, ``\textit{Visualization provides a new perspective to our research. Honestly, we have almost neglected the application of these models beyond the example data. We will consider adding more temporal and spatial factors to the model, and make some visual presentations to demonstrate the results better.''}
 
\par \textbf{Usability and Suggestions.} Generally, experts agreed that different modules in {\name} are easy to use and understand. They also gave some suggestions on our system. First, \textbf{E3} found the {\temporaloverview} a little complex at first sight, but upon mastering it, he discovered its utility. ``\textit{This design is indeed enlightening for analyzing the spatiotemporal changes in traffic hotspots and charging demand. It would be better if users could understand it without any tutorials.}'' Second, he also pointed out, ``\textit{Land use is another important factor for charging station deployment, especially in large cities. It's recommended to display the land use information on the map directly.}'' Further, \textbf{E4} suggested adding more options to deploy new charging stations, such as using the mouse to refine the position of a new charging station.

\section{Discussion and Limitation}
\par \textbf{Improvements on the Experts' Workflow and Prior Studies.} {\name} shows many improvements in CSLP and visual analytics. First, for the decision-making process of CSLP in reality, our system can support the spatiotemporal analysis with more information, solutions generation and post-deployment impact evaluation, which can effectively reduce human effort while enhancing the reliability and rationality of decision-making. Second, {\name} provides a new perspective to CSLP-related research, particularly in the modeling of coupled networks and solving CSLP. Our system utilizes real data and offers a more comprehensive approach to evaluating the deployment results in spatiotemporal dimensions, which has been overlooked in previous studies. Third, although many researchers focus on FLP based on visual analytics, there is less focus on analyzing the impact after the deployment, especially at the network level. {\name} uses a model of the coupled transportation and power networks and implements some novel visual designs to better demonstrate the potential impact at the network level.

\par \textbf{Scalability and Generalizability.} For scalability, {\name} can be easily applied to common scenarios for CSLP. The target area in the case study is approximately 30 $km^2$. According to experts' opinions, the size of this area is sufficient for most charging station deployment tasks. Due to computational complexity and visualization limitations, our system cannot be applied to CSLP with excessively large scopes, such as city-level charging station planning. Additionally, the genetic algorithm also has good scalability. Except for adding new charging stations, as demonstrated in case II, our genetic algorithm offers various deployment options, such as adjusting the number of chargers. Furthermore, new metrics can also be easily added to the objective function. \revised{Regarding generalizability, {\name} demonstrates several strengths. First, the OD data used by {\name} is a common type of travel data that can be easily obtained from sources such as taxis and buses. Second, {\name} can be extended to other scenarios, including analyzing the impact of power grid failures and changes in traffic conditions on charging stations. Third, some of our designs can be applied to other research problems. For instance, the {\temporaloverview}, which provides an overview of traffic hotspots, can be utilized for urban planning and congestion analysis. Similarly, the {\detailview} can be implemented in other systems with two networks, such as the Internet of Vehicles (IoV) and logistics management.}

\par \revised{\textbf{Learning Curve.} Considering experts typically use simple visualizations, we aimed to balance the amount of information presented with usability in our visual design. For example, the {\detailview} was improved from a line chart, and we avoided excessive embedding design in the glyphs of the {\mapview}. Overall, after an approximate 20-minute tutorial, experts could understand the purpose of different modules and the interactions within {\name}. While useful, the {\temporaloverview} and {\mapview} required more time for experts to master. Additionally, researchers familiar with data from the coupled network model tended to grasp the power grid-related visualizations more quickly.}

\par \textbf{Limitation and Future Work.} {\name} has several limitations: \textbf{1) Modeling of the coupled networks.} The model of the coupled networks is based on some basic assumptions and simplifies many factors. Therefore, the results are closer to an ideal condition, which may differ from the real traffic conditions and power grid status. \revised{In the future, we will consider more constraints and use more data collected from the real world.} \textbf{2) Computational cost.} Due to the computational cost of solving the coupled networks model, it is still difficult for our system to handle citywide road networks. Moreover, the computational complexity hinders us from considering the post-deployment impact in the genetic algorithm, since the genetic algorithm may generate dozens to hundreds of solutions in each iteration. Looking ahead, we plan to use some heuristic approaches to solve the model of the coupled networks. \textbf{3) Data range.} Some experts suggested that predicting future charging demand to guide power grid planning would be an interesting problem. However, the increase in EVs may take years to affect charging demand significantly. Existing data in {\name} cannot support such long-term prediction. \revised{In the future, we will consider incorporating data with longer time spans to support the prediction of charging demand.} \revised{\textbf{4) Visualization.} {\name} is designed to assist the decision-making process for CSLP. Some data from the coupled network are omitted as they may not be relevant to this process. We plan to introduce novel visualizations and color schemes to more comprehensively demonstrate the coupling between transportation and power networks.}

\section{Conclusion}
\par We introduce {\name}, a visual analytics system to integrate traffic and power grids for optimal charging station deployment. We identified key metrics and practical needs for CSLP, then analyzed the coupled transportation and power networks using a genetic algorithm to generate deployment solutions. To support decision-making, we employed a community discovery algorithm and innovative visual designs to illustrate spatiotemporal patterns in both networks. Two case studies and expert interviews validated {\name}'s usability and effectiveness. In the future, we aim to include factors like land use and points of interest, and broaden functionality to support diverse deployment options.


\newpage

\acknowledgments{%
This work is supported by grants from the National Natural Science Foundation of China (No. 62302531) and the Science and Technology Planning Project of Guangdong Province (No. 2023B1212060029).
}

\balance
\bibliographystyle{abbrv-doi-hyperref}

\bibliography{template}

\appendix 

\end{document}